\documentclass[aps,prl,groupedaddress,twocolumn,superscriptaddress,amsfonts,amssymb,amsmath,showpacs,floatfix,preprintnumbers]{revtex4-2}

\usepackage[utf8]{inputenc}
\usepackage{amsmath}
\usepackage{xcolor}
\usepackage{soul}
\usepackage{graphicx}
\usepackage{tikz}
\usepackage{float}
\usepackage{amsfonts}
\usepackage{amssymb}
\usepackage{dsfont}
\usepackage[toc,page]{appendix}
\usepackage[top=1in,bottom=1in,right=1in,left=1in]{geometry}
\usepackage{amsthm}
\usepackage{subfiles}
\usepackage{braket}
\usepackage{listings}
\usepackage{slashed}
\usepackage[obeyspaces]{xurl}
\usepackage{hyperref}
\usepackage{booktabs}
\usetikzlibrary{decorations.markings}
\usetikzlibrary{decorations.pathmorphing}

\definecolor{crim}{RGB}{192,64,98}
\NewDocumentCommand{\codeword}{v}{\texttt{{#1}}}
\DeclareMathAlphabet{\mathpzc}{OT1}{pzc}{m}{it}
\usetikzlibrary{positioning}
\setulcolor{red}
\newcommand{\ztwo}{\frac{z}{2}}

\hypersetup{colorlinks=true, linkcolor=blue, citecolor=blue, urlcolor=blue}
\begin{document}

\title{Three-dimensional parton structure in QCD with $CP$-violation}

\author{Alexandru-Mihail Sturzu }
\email[e-mail: ]{amsturzu@wm.edu}
\affiliation{Physics Department, William \& Mary, Williamsburg, VA 23187, U.S.A.} 
\author{Joe Karpie}
\email[e-mail: ]{jkarpie@jlab.org}
\affiliation{Thomas Jefferson National Accelerator Facility, Newport News, VA 23606, U.S.A.}
\preprint{JLAB-THY-26-5034}
\date{\today}

\begin{abstract}
The Energy-Momentum tensor encodes the mechanical properties of a system such as the internal distributions energy, momentum, forces, and particle fluxes. The QCD energy momentum tensor respects $CP$ symmetry even when QCD is modified with a $CP$-violating $\theta$-term. In this theory, there are 5 unique generalized form factors in the Lorentz decomposition of the total Energy Momentum tensor and 3 more which only vanish when summing over all flavors. We introduce two new light-cone GPDs whose Mellin moments give the electromagnetic and gravitational form factors. These generalize the observables for QCD-based $CP$ violation such as the neutron electric dipole moment to three dimensional light cone structure.
\end{abstract}
\maketitle
\section{Introduction}
%
\par
The energy-momentum tensor (EMT) of a system, such as a proton, encodes how the internal energy, pressure, and other mechanical properties are distributed.
These features provide insight into how the constituents, called partons, conspire to form the bulk properties.
For hadronic systems, the EMT provides distributions of the mass, spin, internal forces and flux.
\par
The off-forward matrix elements of the EMT are parameterized by the generalized form factors (GFFs).
In an analogous way to how the form factors of the electromagnetic current detail properties such as the charge and magnetic moment distribution, the form factors of the EMT detail the mechanical properties and distribution of forces.
These form factors are of great interest both theoretically and phenomenologically as they provide a window into the mechanical and internal properties of hadrons.
\par 
In the standard QCD Lagrangian, the theory is $CP$‑invariant and the nucleon matrix elements of the total, conserved EMT $T^{\mu\nu}$ can be decomposed into three, non-zero, independent gravitational form factors. Specifically, for the Belinfante improved EMT~\cite{belinfante1940current,rosenfeld1940energy}, the matrix elements decompose into the GFFs $A(t)$, $B(t)$ and $C(t)$. Some conventions introduce an additional trace term $\bar{C}(t)$; such that in the sum of quark and gluon contributions $\bar{C}(t)=0$. These nucleon GFFs have been targets of recent Lattice QCD calculations.~\cite{Hackett:2023rif,Bhattacharya:2023ays,HadStruc:2024rix,Bhattacharya:2024wtg,Alexandrou:2026tjs}

When a $CP$-violating $\theta$-term is added to the QCD action, the EMT operator is unchanged, but the matrix elements are affected because of the effects on the hadronic states.
Therefore nothing forbids $CP$‑odd Dirac structures in its matrix elements.
As a result, additional pseudoscalar or axial structures can appear, leading to new gravitational form factors.
Our goal is to perform this decomposition explicitly.
\par
There has been significant interest in the study of $CP$-violating effects in QCD.
The absence of observed $CP$ violation in strong interactions, despite the theoretical allowance for a $\theta$-term, is the famous strong $CP$ problem. 
See Ref.~\cite{DiLuzio:2020wdo} for a review on this subject.
Most sensitive of experimental probes has been the neutron electric dipole moment (nEDM), which constrains the $\theta$ parameter to be extremely small, $|\theta| \lesssim 10^{-10}$ \cite{Abel:2020pzs}.
There have been several theoretical proposals to explain this smallness, including the Peccei-Quinn mechanism \cite{Peccei:1977hh, Peccei:1977ur, Peccei:2006as} and related axion models \cite{DiLuzio:2020wdo}.
However, a lack of experimental signal for axions \cite{ParticleDataGroup:2024cfk} has motivated continued interest in the direct study of $CP$-violating effects in QCD, including the effects of a non-zero $\theta$ on hadronic properties.
This is the motivation and context for exploring the EMT decomposition in the presence of a $\theta$-term.
\par
Certain GFFs are related to the Mellin moments of Generalized Parton Distributions (GPDs). GPDs encode the internal structure of the hadron in terms of the constituent's longitudinal momentum fraction $x$, the longitudinal momentum transfer encoded in skewness $\xi$ and the total momentum transfer $t$. Specific integrals of the GPDs give the Impact Parameter Distributions (IPDs) which provide a 3-dimensional image of the hadron in terms of the parton's transverse separation or impact parameter and either the parton's momentum fraction $x$~\cite{Burkardt:2002ks} or the parton's interaction time or Ioffe time $\nu$~\cite{Edwards:2026egu}.

GPDs can be accessed experimentally through processes such as deeply virtual Compton scattering (DVCS)~\cite{Muller:1994ses,Ji:1996nm,Radyushkin:1996nd}, deeply virtual meson production (DVMP)~\cite{Radyushkin:1996ru,Collins:1996fb} and constrained by single diffractive hard exclusive processes (SDHEP)~\cite{Qiu:2022pla}, which are 
measured, or planned to be measured, at
CEBAF~\cite{Dudek:2012vr,Accardi:2023chb}, J-PARC~\cite{Sawada2016}, SPS~\cite{Gautheron:1265628,Adams:2676885} and the EIC~\cite{Accardi:2012qut}. GPDs are also the target of lattice QCD calculations~\cite{Chen:2019lcm,Alexandrou:2020zbe,Lin:2020rxa,Bhattacharya:2022aob,Bhattacharya:2023nmv,Lin:2023gxz,Bhattacharya:2024qpp,Ding:2024saz,Bhattacharya:2025yba,Chu:2025kew,Dutrieux:2026grg}.
Phenomenological analysis of experiments and other constraints, some even including lattice QCD data, have begun to extract GPDs~\cite{Kumericki:2016ehc,Kumericki:2017gdc,Guo:2025muf,Xu:2026lko,Mezrag:2026wcf}.
Similar questions about the influence of a non-zero topological term are relevant in this sector as well. 
\par
In this work we discuss prospects for using hadronic processes and structure studies to probe for a non-zero $\theta$ term.
We do this by considering the Lorentz decomposition of matrix elements of the energy-momentum tensor. 
When enumerating the GFFs in the presence of a non-zero $\theta$-term, we find five structures that survive after imposing the on-shell Dirac and Gordon identities, hermiticity, $PT$ invariance, and energy-momentum conservation.  
This is one fewer than the decompositions of \cite{Kobzarev:1962wt} and \cite{Ng:1993vh} the former of which considered gravitational interactions with fermions; the latter extended their work to neutrinos.
Note, we neglect gauge-fixing and ghost contributions since we are ultimately interested in the Lorentz decomposition of the matrix elements of the EMT operator between physical hadronic states.
\par
Other studies of parton structure of QCD with a $\theta$-term have introduced parity-violating fragmentation functions~\cite{Kang:2010qx} and parton distribution functions (PDF)~\cite{Seng:2018wwp,Yang:2019hxu} and investigated their contributions to inclusive deep-inelastic scattering \cite{Bacchetta:2023hlw}. These are different from what will be studied below in that they are studying matrix elements of operators which are usually dropped due to parity, while the Lorentz decompositions of the GFFs and GPDs presented below are the matrix elements of standard $CP$-even operators and we are asking what $CP$-odd contributions can be generated when $CP$ is broken.
We extend these investigations to the non-local lightcone correlators introducing two new unpolarized and two new helicity GPDs which appear in the presence of $CP$-violation. We identify the relationships between the new GFFs and GPDs. There will also be GPDs generalizations of the parity-violating PDF as well to be investigated in future work.
\par
The outline of this work is as follows.
In Section~\ref{EMTwithTheta} we present the Lorentz decomposition of the EMT for nonzero $\theta$. 
Section~\ref{sec:gpds}  extends these considerations to the matrix elements defining GPDs and derive their relationship to the GFFs.
Finally, Sect.~\ref{sec:conclusions} summarizes our findings and suggests future directions.
\section{EMT in $\theta$-QCD} \label{EMTwithTheta}
The QCD action permits an additional term built from the gauge field-strength tensor and its dual,
\begin{align} \label{eq:thetaTerm}
    \mathcal{L}_{\mathrm{\theta QCD}} = \overline{\psi}\left( i \slashed{D} - m \right)\psi - \frac{1}{4} &G^{\mu \nu}_a G^a_{\mu \nu} \nonumber \\
    - \frac{\theta g^2}{32\pi^2} &G^{\mu \nu}_a \tilde{G}^a_{\mu \nu},
\end{align}
Here, $G^a_{\mu\nu} = \partial_\mu A^a_\nu - \partial_\nu A^a_\mu + g f^{abc} A^b_\mu A^c_\nu$  and $\tilde G^{a\mu\nu} = \frac{1}{2} \epsilon^{\mu \nu \rho \sigma} G^a_{\rho \sigma}$ are the field strength tensor and its dual.
%
%
Including a $CP$-violating $\theta$ term in the QCD Lagrangian does not modify the local operator form of the EMT.
It is the vacuum and hadronic states that are altered. 
Thus, the $CP$-odd structures result from these altered states, not from a change to the EMT operator itself.
\par
For this work, we use the Belinfante-improved EMT, which is symmetric and gauge-invariant
\begin{align}
    T^{\mu \nu} = \frac{i}{4}\overline{\psi}\left( \gamma^{\mu}\overleftrightarrow{D}^{\nu} + \gamma^{\nu}\overleftrightarrow{D}^{\mu} \right)  \psi  - &G^{\mu \alpha}_{a}G^{\nu a}_{\alpha} \nonumber \\
    + \frac{1}{4} g^{\mu \nu} &G^{\rho \sigma}_a G^a_{\rho \sigma}
\end{align}
Notice that the field strength tensor remains unchanged for nonzero $\theta$.
The hadronic GFFs are obtained by decomposing the off-forward matrix elements of the EMT into the allowed Lorentz covariant structures,
\begin{align}
    \langle p',\lambda'|\,T^{\mu\nu}\,|p,\lambda\rangle &= \bar{u}(p',\lambda') \Gamma^{\mu\nu}(P,\Delta) u(p,\lambda) \nonumber \\
    &= \sum_i \mathcal{A}_i (t) \mathcal{T}^{\mu \nu}_i(P, \Delta,\lambda,\lambda').
\end{align}
where $P=\frac{p'+p}{2}$, and $\Delta=p'-p, \; \Delta^2=t$ are the total momentum and momentum transfer variables and $\lambda$ and $\lambda'$ are the nucleon helicities.
The Dirac spinors $u(p,\lambda)$ and $\bar{u}(p',\lambda')$ correspond to the initial and final spin‑$1/2$ hadron states, respectively.  
The possible Lorentz structures, $\mathcal{T}^{\mu \nu}_i$, are constrained by Lorentz invariance, hermiticity, and discrete symmetries such as parity and time reversal. 
These constraints are used to reduce the number of independent terms in the decomposition.
\par
The set of symmetric rank‑two tensors, $\mathcal{T}^{\mu \nu}_i$, are constructed from a basis of Lorentz and Dirac structures multiplying the kinematic variables $P$ and $\Delta$.
The only available Lorentz and Dirac structures are,
\begin{align*}
\text{Lorentz structures} &:\quad g^{\mu\nu},\ \epsilon^{\alpha\beta\mu\nu},\ P^\mu,\ \Delta^\mu, \\
\text{Dirac structures} &:\quad \mathbb{I},\ \gamma^\mu,\ \sigma^{\mu\nu},\ \gamma^5,\ \gamma^5\gamma^\mu,\ \gamma^5\sigma^{\mu\nu}.
\end{align*}
Notice the doubling of available Dirac structures from $CP$-symmetry breaking.
As detailed in the Appendix \ref{App:LaD_removal}, combinations and entire structures, like the Levi-Civita tensor, can be immediately removed from this overcomplete list. 
After removing such terms, an explicit decomposition is given in Eq.~\eqref{eq:overcomplete}.
This decomposition is then reduced further using the Dirac equation, Gordon identities, hermiticity, and discrete symmetry constraints. 
The total number of independent factors can be reduced to eight terms 
\begin{align} \label{eq:emt-flavour-decomposition}
\Gamma_a^{\mu\nu} &= A_a(t) \gamma^{\{\mu}P^{\nu\}} + B_a(t) \frac{P^{\{\mu}i\sigma^{\nu\}\rho}\Delta_\rho}{2M} \nonumber\\
&+ C_a(t) \frac{\Delta^\mu\Delta^\nu-g^{\mu\nu}t}{M} + M g^{\mu\nu} \bar C_a(t) \nonumber\\
&+  B_{5,a}(t) \frac{i \gamma_5}{M} P^{\mu}P^{\nu} + C_{5,a}(t) \frac{i\gamma_5}{M} \left(\Delta^\mu\Delta^\nu-g^{\mu\nu}t\right)  \nonumber\\
&+ g^{\mu\nu} iM\gamma_5\bar C_{5,a}(t) + i \bar F_{5,a}(t)\, \gamma_5 \gamma^{\{\mu} \Delta^{\nu\}}.
\end{align}
\par
The conservation of the EMT, $\Delta_{\mu} T^{\mu \nu}(t) = 0$, reduces this set further.
The conservation equation manifests as three distinct constraints on the set of allowed structures for the total EMT.
The additional constraints come from the linear independence of the tensor structures that remain after contracting with $\Delta_{\mu}$.
Ultimately, in contrast to \cite{Kobzarev:1962wt,Ng:1993vh}, we show that the EMT in $\theta$-QCD has only 5 linearly independent terms in its decomposition. 
In the appendix we provide two different counts of the independent structures, including a complete derivation of the remaining GFFs.
The subscript $a$ specifies quark flavor or gluon choice.
Conservation of the EMT requires $\sum_a \bar C_a = \sum_a\bar C_{5, a} = \sum_a \bar{F}_{5,a} = 0$, leaving 5 structures.
Three of these are the standard parity-even form factors, $A(t)$, $B(t)$ and $C(t)$, while the remaining two are $CP$-odd form factors, $B_5(t)$ and $C_5(t)$. 
\par
Although the $\theta$‑term breaks $CP$, $CPT$ invariance remains intact. 
Therefore, the full matrix element of the EMT must satisfy $CPT$ invariance, which constrains the real and imaginary parts of the form factors. 
In particular, hermiticity implies that all GFFs are real functions for spacelike momentum transfer. 
\section{GPDs in $\theta$-QCD}\label{sec:gpds}
Generalized parton distributions appear in the factorization of DVCS, DVMP, and other SDHEPs. They encode the structure of the hadron in terms of its longitudinal momentum fraction $x$, the skewness, which is the longitudinal momentum fraction transferred $\xi$, and the total momentum transfer $t$. They are given by the Fourier transform of the generalized Ioffe time distributions (GITD). Those in turn are the Lorentz invariant functions in the decomposition of the matrix elements
\begin{equation}
I^{q,g}(z, p', p, \lambda', \lambda) = \langle p', \lambda' | O^{q,g}(z) | p, \lambda\rangle
\end{equation}
 where for quarks
 \begin{equation} 
 O^q(z) = \bar{q}(-\ztwo)\slashed{z} W(-\ztwo,\ztwo)q(\ztwo)
 \end{equation}
 and, for gluons,
\begin{equation}
O^g(z) = z_\mu z_\nu G_a^{\mu i}(-\ztwo) W(-\ztwo,\ztwo)G_{i}^{\nu a}(\ztwo)\,.
\end{equation}
where $i$ are summed transver indices.
 Here $z$ is in the direction of the light cone defined by the spatial direction of the average momentum $P$, conventionally, such that $z^+=0$.

Both of these matrix elements have a Lorentz decomposition into 4 GITDs
\begin{eqnarray}
    I^{q,g}(z, p', p, \lambda', \lambda) =& \bar{u}(p',\lambda')\slashed{z} u(p,\lambda) \mathcal{H}^{q,g}(\nu, \bar{\nu},t)  \nonumber\\
    + & \bar{u}(p',\lambda') \frac{i \sigma^{zq}}{2m} u(p,\lambda) \mathcal{E}^{q,g}(\nu, \bar{\nu},t) \nonumber\\
    + & i\bar{u}(p',\lambda')\slashed{z} \gamma_5 u(p,\lambda) \mathcal{H}_5^{q,g}(\nu, \bar{\nu},t)  \nonumber\\
    + & \bar{u}(p',\lambda') \frac{\sigma^{zq} \gamma_5}{2m} u(p,\lambda) \mathcal{E}_5^{q,g}(\nu, \bar{\nu},t)\,. \nonumber\\\label{eq:gitds}
\end{eqnarray}
Here we introduce two new GITDs that appear in the presence of a $\theta$ term, $\mathcal{H}_5$ and $\mathcal{E}_5$, from now on dropping the parton's flavor label.

All GITDs are defined in terms of Lorentz invariants $\nu=P\cdot z$, $\bar{\nu} = -\Delta\cdot z /2$, and $t=\Delta^2$. The Lorentz invariant definition of skewness is given by $\xi=\bar{\nu}/\nu$ which gives the traditional ratio for light cone $z$. At fixed $z$, $p'$, and $p$, all possible combinations of $\lambda,\lambda'$ create a $2\times2$ complex matrix. Such a matrix can be completely parameterized by 4 independent complex GITDs, so no more terms are needed. Their values at $\nu=0=\bar{\nu}$ are given by local $z_-=0$ matrix elements. $\mathcal{H}(0,0,t)$ and $\mathcal{E}(0,0,t)$ give the Dirac EM form factors $F_1$ and $F_2$. $\mathcal{H}_5(0,0,t)$ gives the parity-violating anapole form factor $F_A$ and $\mathcal{E}_5(0,0,t)$ is the $F_3(t)$ form factor related to neutron electric dipole moment.

The GPDs themselves are given by the Fourier transform of $\nu$ at fixed skewness $\bar{\nu} =\xi\nu$
\begin{equation}
    H_5(x,\xi,t) = \int_{-\infty}^\infty \frac{d\nu}{2\pi} e^{i\nu x}  \mathcal{H}_5(\nu,\xi\nu,t) \,.\label{eq:gpd_def}
\end{equation}
and similarly for $H$, $E$, and $E_5$ from their respective GITDs. Following Refs.~\cite{Radyushkin:1996nd,Radyushkin:2023ref}, Double Distributions
\begin{equation}
    h_5(\beta,\alpha,t) = \int_{-\infty}^\infty \frac{d\nu}{2\pi}e^{i\nu \beta} \int_{-\infty}^\infty \frac{d\bar\nu}{2\pi}e^{i\bar\nu \alpha} \mathcal{H}_5(\nu,\bar{\nu},t)\,,
\end{equation}
are defined and their relationship to the GPDs explain the standard polynomiality property of GPD Mellin moments.
 Finally, there are impact parameter distributions in momentum fraction ~\cite{Burkardt:2002ks} or Ioffe time~\cite{Edwards:2026egu} spaces calculated from the Fourier transform of transverse $\Delta_T$ at $\xi=0$
 \begin{equation}
 h_5(\nu,b_T) = \int \frac{d^2\Delta_T}{(2\pi)^2}    e^{-i\Delta_T\cdot b_T} \mathcal{H}_5(\nu,0,-\Delta_T^2)\,,
 \end{equation}
 for example.

 Finally, the first and second Mellin moments in $x$ are related to the EM form factors and the Gravitational form factors. Table.~\ref{tab:gpd_to_mellin} shows the relationship between the specific GPDs, Dirac EM FFs and the GFFs in Sec.~\ref{EMTwithTheta}, which can be found by contracting the vector current or EMT with $z_\mu$.
 \begin{table}[]
     \centering
     \begin{tabular}{c|c| c}
         GPD\,\,  &  \,EM FF ($n=1$)\, & \,GFF  ($n=2$)\, \\\hline
          $H$ &  $F_1$ & $A+4\xi^2C$ \\
          $E$ &  $F_2$ & $B - 4 \xi^2 C$ \\
          $H_5$ &  $F_A$ & $ 2\xi \bar{F}_5$ \\
          $E_5$ &  $F_3$ & $B_5 + 4\xi^2 C_5$ \\
     \end{tabular}
     \caption{Table of GPDs and related Mellin moments.}
     \label{tab:gpd_to_mellin}
 \end{table}

The symmetry properties of the spinor bilinears and of the matrix element will dictate the $(\nu, x,\xi)$ parities of the GITD and GPD. Consider instead matrix elements of the hermitian eigen-operators $O_\pm(z) = \frac12 (O(z) \mp O(-z))$ which have the properties $O_\pm^\dagger(z)=\pm O_\pm(z)$ and $O_\pm(-z)=\mp O_\pm(z)$. Their matrix elements expose even/odd behavior in the GITDs' invariants. The factors of $i$ prescribed in the Lorentz decomposition Eq.~\eqref{eq:gitds} are such that the real components of the GITDs are even and the imaginary components components are odd under $(\nu,\bar{\nu})\to(-\nu,-\bar{\nu})$. From Eq.~\eqref{eq:gpd_def} one can also see the real components are the cosine transforms of $H_5(x,\xi,t)+H_5(-x,\xi,t)$ and the imaginary components are the sine transforms of $H_5(x,\xi,t)-H_5(-x,\xi,t)$. 

Consider the impact of hermitian conjugation, parity and time reversal as done in Ref~\cite{HadStruc:2024rix}. These combine to the transformation of $\xi\to-\xi$ and $\lambda_{i,f}\to -\lambda_{i,f}$. Using spinor identities to invert the $\lambda_{i,f}$ sign again, one finds $H$, $E$, and $E_5$ to be even under $\xi\to -\xi$ while $H_5$ is odd. Polynomiality forces the Mellin moments to be polynomials of $\xi$ whose order depends on the moment's power. The Mellin moments are exposed in the Taylor expansion of any generic GITD
\begin{eqnarray}
\mathcal{G}(\nu,\xi\nu,t) =& \sum_{n=1}^\infty \frac{(-i\nu)^{n-1}}{(n-1)!} \int_{-1}^1 dx x^{n-1} G(x,\xi,t) \nonumber\\ =& \sum_{n=1}^\infty \frac{(-i\nu)^{n-1}}{(n-1)!} \sum_{k=0}^{n} \xi^k G_{nk}(t)
\end{eqnarray}
and the parity of $\nu$ and $\xi$ govern which $G_{nk}(t)$ vanish where $\xi^n$ term is from the $D$ term explicitly.
For $H_5$ which is odd in $\xi$, the first moment, the anapole form factor $F_A$, vanishes
\begin{equation}
    F_A(t) = \int H_5(x,\xi,t) dx = \mathcal{H}_5(0,0,t)=0 \,.
\end{equation} 
Since $\bar{F}_5$ vanishes when summed over all flavors, the sum of the second Mellin moments of $H_5$ also vanishes.

Similar to the unpolarized GITDs are the helicity distributions from the matrix elements $\tilde{I}^{q,g}$ of the operators
 \begin{equation} 
 \tilde{O}^q(z) = \bar{q}(-\ztwo)\slashed{z} \gamma_5 W(-\ztwo,\ztwo)q(\ztwo)
 \end{equation}
 and
\begin{equation}
\tilde{O}^g(z) = z_\mu z_\nu G_a^{\mu i}(-\ztwo) W(-\ztwo,\ztwo)\tilde{G}_i^{\nu a}(\ztwo)\,.
\end{equation}
These will have the Lorentz decomposition
\begin{eqnarray}
    \tilde{I}^{q,g}(z, p', p, \lambda', \lambda) =& 
     \bar{u}(p',\lambda')\slashed{z} \gamma_5 u(p,\lambda) \tilde{\mathcal{H}}^{q,g}(\nu, \bar{\nu},t)  \nonumber\\
    + & \bar{u}(p',\lambda') \frac{i\sigma^{zq} \gamma_5}{2m} u(p,\lambda) \tilde{\mathcal{E}}^{q,g}(\nu, \bar{\nu},t)\nonumber\\
    + &i\bar{u}(p',\lambda')\slashed{z} u(p,\lambda) \tilde{\mathcal{H}}_5^{q,g}(\nu, \bar{\nu},t)  \nonumber\\
    + & \bar{u}(p',\lambda') \frac{\sigma^{zq}}{2m} u(p,\lambda) \tilde{\mathcal{E}}_5^{q,g}(\nu, \bar{\nu},t) \,. \nonumber\\\label{eq:hel_gitds}
\end{eqnarray}
into the 4 helicity GITDs. Similar definitions occur for the related GPDs, DDs, and IPDs. The Mellin moments of the helicity GPDs will be related to the Lorentz decomposition of the axial vector current and other generalized form factors for higher moments. Quark transversity GITDs/GPDS as well as new GITDs/GPDs generalizing the $CP$-odd PDFs of Ref.~\cite{Seng:2018wwp,Yang:2019hxu} also can be defined by their own Lorentz decompositions.

\section{Conclusions and outlook}\label{sec:conclusions}
%
This work considers the energy-momentum tensor and generalized parton distributions of the nucleon in QCD with a $\theta$ term. The $CP$ violations of the Lagrangian do not translate to $CP$ violation of the EMT operator itself. Still, changes in the vacuum and hadronic state create matrix elements that violate $CP$. An overcomplete list of Lorentz structures is given and 8 remain after applying equations of motion, and symmetry constraints. Conservation of the EMT reduces to 5 GFFs for the sum of flavors. Similarly, two new unpolarized GPDs and two new helicity GPDs appear in the decomposition of the standard light cone matrix elements. 

The derivation may be extended to higher‑spin hadrons, transition form factors, GPD extensions of the $CP$-odd PDFs of Ref.~\cite{Seng:2018wwp,Yang:2019hxu}, and the uncontracted GPD operators used for lattice~\cite{Bhattacharya:2022aob,Schoenleber:2024auy}, where the number of independent $CP$‑odd structures may differ. Systematic methods such as in Ref.~\cite{Ji:2000id} applied in App.~\ref{App:jpc_count} will be necessary for finding the minimal basis. Ref~\cite{Sun:2026wqa} demonstrates an approach for determining GFFs applied to bosons and fermions of many different spins. 

Quantitative estimates of new GFFs and GPDs, as well as the $\theta$-QCD corrections to traditional GFFs and GPDs, could be obtained using chiral effective theory, holographic QCD, QCD sum rules or lattice simulations that incorporate a small $\theta$ term or imaginary $\theta$ methods. Perturbatively approximating the $\theta$ term for lattice GFF and GPD calculations is possible through recycling the majority of numerical work already performed, though statistical quality may be lacking without additional computation. The potential interplay with axion dynamics and implications for nucleon structure are also worthy of exploration. 

In principle, the GPDs introduced from $CP$ violation can be measure via forbidden modes within the scatterings angles of DVCS or other SDHEPs. Consider the SDHEP frame~\cite{Qiu:2024reu} which describe azimuthal angles based upon the plane of the diffracting hadrons' momenta and the momenta plane of the particles leaving the hard scattering. In the case of DVCS with $CP$ conservation, scattering unpolarized nucleon and electron leads to $\cos\phi$ harmonics while longitudinally polarized electrons off unpolarized nucleons leads to $\sin\phi$ harmonics. Parity violation could be measured in the forbidden $\sin\phi$ and $\cos\phi$ modes respectively. The weak sector $P$-violation could be a substantial background over these signals which would need to be handled due to $\theta$'s expected small size. A cleverly chosen and precisely measured SDHEPs, or asymmetries of SDHEPs such as comparing electron scattering to positron scattering at C$e^+$BAF~\cite{Grames:2023lxn}, could provide an alternative, though potentially difficult, discovery mechanism for $\theta$-QCD. Future analysis is needed to find ideal observables which have sufficiently suppressed weak sector $CP$ violation, so that $\theta$-QCD can be meaningfully constrained to the same extent of neutron EDM measurements.

\section{Acknowledgments}
The authors gratefully thank Adam Freese and Jianwei Qiu for productive discussions.

This material is based upon work supported by the U.S. Department of Energy, Office of Science, Office of Nuclear Physics under Contract No. 89243126CSC000213. A.M.S. is supported in part by U.S. DOE Grant \#DE-SC0025908. This work has benefited from the collaboration enabled by the Quark-Gluon Tomography (QGT) Topical Collaboration, U.S.~DOE Award \mbox{\#DE-SC0023646}. 

\appendix \label{appendix}
\section{Lorentz and Dirac structure reductions} \label{App:LaD_removal}
From the available Lorentz and Dirac structures, $\Gamma^{\mu \nu}$ will contain terms of the form
\begin{align}
\mathcal T_{\mathbf n,A,c}^{\mu\nu} = \left[ P^{\otimes n_P} \otimes \Delta^{\otimes n_\Delta} \otimes g^{\otimes n_g} \otimes \epsilon^{\otimes n_\epsilon} \otimes D_A \right]_{c}^{\{\mu\nu\}}.
\end{align}
Where the Dirac factor, $D_A$, is kept general since arbitrary products of gamma matrices can be reduced to a linear combination of $\{ \mathds{1}, \gamma_5, \gamma^{\mu}, \gamma_5 \gamma^{\mu}, \sigma^{\mu \nu} \}$.
The exponents, $\{n^P, n_{\Delta}, n_g, n_{\epsilon} \}$, count the number of occurrence of each structure.
The $c$ subscript labels a particular contraction of all the indices, that necessarily leaves $\mu$ and $\nu$ free.
\par
The Levi-Civita tensor can be removed from consideration completely.
To see this, first note that any product containing two Levi--Civita tensors can be reduced to products of metric tensors through the generalized Kronecker delta, $\epsilon^{\mu\nu\alpha\beta} \epsilon_{\rho\sigma\lambda\kappa} = -\delta^{\mu\nu\alpha\beta}_{\rho\sigma\lambda\kappa}$.
Therefore, terms containing more than one Levi--Civita tensor generate no new Lorentz structures, and we need only argue away terms with a single Levi-Civita tensor.
Additionally, such terms have at most 1 free index. Both free indices cannot come from the anti-symmetric tensor because we restrict to a symmetric EMT.
\par
The remaining terms can be classified according to the Dirac factor $D_A$.
For terms containing the Dirac scalars, $\mathds{1}$ and $\gamma_5$, the Levi-Civita tensor must contract with at least three kinematic vectors.
Since there are only two kinematic vectors, any choice will disappear by antisymmetry.
The Levi-Civita tensor can be removed from terms containing tensor Dirac structures by use of the duality relation
\begin{align}
\sigma^{\rho\sigma} = \frac{i}{2}\epsilon^{\rho\sigma\alpha\beta} \sigma_{\alpha\beta}\gamma_5,
\end{align}
along with the genrealized Kronecker expression introduced earlier.
For a vector or axial-vector Dirac factor, nonzero contributions must contract the vector's index with the Levi--Civita tensor otherwise duplicated kinematic vectors are contracted and thereby 0 by antisymmetry.
The other two indices must be contracted with $P$ and $\Delta$.
Then, by use of the identity
\begin{align}
    \epsilon^{\rho \kappa \alpha \beta} \gamma_{\beta} = -i\gamma^{\kappa}\gamma^{\alpha}\gamma^{\rho}\gamma_5 + ig^{\kappa \alpha} \gamma^{\rho} \gamma^5 &- ig^{\kappa \rho} \gamma^{\alpha} \gamma^5 \nonumber \\
    &+ ig^{\alpha \rho} \gamma^{\kappa} \gamma^5,
\end{align}
such terms can be reduced to products of metric tensors and $\gamma$-matrices.
For the terms that contain three $\gamma$-matrices, this results in terms like $\slashed{P}\slashed{\Delta}$.
Such terms can be preemptively removed since when they are placed between spinors they can be reduced to single $\gamma$-terms by means of the Dirac equation after some reordering.
Therefore, any term containing the Levi-Civita tensor can be recast into terms not containing the tensor or whose matrix elements evaluate to 0. 
\par
The Dirac equation for the total and momentum transfer are $\bar{u}(p')\,\slashed{P}\,u(p) = m\,\bar{u}(p')u(p)$ and $\bar{u}(p')\,\slashed{\Delta}\,u(p) = 0$, where for convenience here and in the following we drop the helicity argument.
Therefore, terms proportional to $\slashed{P}$ and $\slashed{\Delta}$ can be absorbed into the coefficient functions $a_i$ or disappear altogether.
The tensor structures that remain after removing combinations involving $\epsilon^{\alpha \beta \gamma \delta}$ are
\begin{widetext}
\begin{align}
\Gamma^{\mu\nu} ={}& a_1(t)\,g^{\mu\nu} + a_2(t)\,P^\mu P^\nu + a_3(t)\,\Delta^\mu \Delta^\nu + a_4(t)\,P^{\{\mu}\Delta^{\nu\}} + a_5(t)\,\gamma^{\{\mu} P^{\nu\}} + a_6(t)\,\gamma^{\{\mu} \Delta^{\nu\}}\nonumber\\[1ex]
 + &\, a_7(t)\,P^{\{\mu}\sigma^{\nu\}\alpha}\Delta_\alpha + a_8(t)\,\Delta^{\{\mu}\sigma^{\nu\}\alpha}P_\alpha + a_9(t)\,P^{\{\mu}\sigma^{\nu\}\alpha}P_\alpha + a_{10}(t)\,\Delta^{\{\mu}\sigma^{\nu\}\alpha}\Delta_\alpha  + a_{11}(t)\,\gamma^5 g^{\mu\nu} \nonumber\\[1ex]
+&\, a_{12}(t)\,\gamma^5 P^\mu P^\nu + a_{13}(t)\,\gamma^5 \Delta^\mu \Delta^\nu + a_{14}(t)\,\gamma^5 P^{\{\mu}\Delta^{\nu\}} + a_{15}(t)\,\gamma^5\gamma^{\{\mu} P^{\nu\}}    + a_{16}(t)\,\gamma^5\gamma^{\{\mu} \Delta^{\nu\}}\nonumber\\[1ex]
+&\, a_{17}(t)\,\gamma^5 P^{\{\mu}\sigma^{\nu\}\alpha}\Delta_\alpha + a_{18}(t)\,\gamma^5 \Delta^{\{\mu}\sigma^{\nu\}\alpha}P_\alpha  + a_{19}(t)\,\gamma^5 P^{\{\mu}\sigma^{\nu\}\alpha}P_\alpha + a_{20}(t)\,\gamma^5 \Delta^{\{\mu}\sigma^{\nu\}\alpha}\Delta_\alpha. \label{eq:overcomplete}
\end{align}
\end{widetext}
%
Notice the terms that appear alongside $a_i$ for $i\geq 11$ are the P-violating terms.
\par
Using the Gordon identity, $\bar{u}(p')\gamma^{\mu}u(p) = \bar{u}(p')\left[\frac{P^\mu}{m} + \frac{i}{2m}\sigma^{\mu \Delta} \right] u(p)$ several of these terms can be further reduced. 
First, consider
\begin{align}
\bar{u}(p')P^{\{\mu}\sigma^{\nu\}\Delta} u(p) = &-2mi\,\bar{u}(p')P^{\{\mu} \gamma^{\}\nu}u(p) \nonumber \\
&+2i\,P^\mu P^\nu\,\bar{u}(p')u(p).
\end{align}
This shows that the tensor structures multiplying $a_2$ are related to those multiplying $a_5$ and $a_7$. 
A similar argument can be made to relate $a_4$ to the terms $a_6$ and $a_{10}$.
\par
A slightly more involved argument is needed to show that the tensor structures multiplying $a_3$ and $a_8$ are related. 
The term multiplying $a_8$ can be simplified after expanding out $\sigma^{\nu \} P}$ and using the Dirac equation
\begin{align}
\Delta^{\{\mu}\bar{u}(p')\sigma^{\nu\} P} u(p) &= \frac{i\Delta^{\{\mu}}{4}\, \bar{u}(p') \Big[ \gamma^{\nu\}}\slashed{p}'- \slashed{p}\gamma^{\nu\}}\Big]u(p) \nonumber \\
&= \frac{i}{2}\Delta^\mu \Delta^\nu \,\bar{u}(p')u(p).
\end{align}
The second equality comes after rewriting the remaining terms using $\gamma^\nu \slashed{p}' = \{\gamma^\nu,\slashed{p}'\} - \slashed{p}'\gamma^\nu$ then simplifying with the Clifford algebra identity $\{\gamma^\mu,\gamma^\nu\} = 2g^{\mu\nu}$ and Dirac equation.
Therefore, $a_8$ is not independent of $a_3$ and can be removed from the remaining set of terms.
A similar relationship holds between the tensor structures of $a_4$ and $a_9$ after replacing $\Delta^\mu$ with $P^\mu$. 
\par
A symmetric set of arguments holds for the corresponding axial structures.
By rewriting $\sigma^{\mu\nu} = i\left( \gamma^\mu \gamma^\nu - \eta^{\mu\nu} \right)$ and $\sigma^{\mu\nu} = i\left( \eta^{\mu\nu} - \gamma^\nu \gamma^\mu \right)$ the axial Gordon identity 
\begin{align} \label{eq:axgor}
\bar{u}(p')\,\gamma^5\left[\frac{P^\mu}{m} + \frac{i}{2m}\sigma^{\mu\alpha}\Delta_\alpha\right]u(p) = 0
\end{align}
is easily derived, and notably simpler than the standard case.
\par
Proceeding in a similar way as above by focusing on the tensor structures multiplying $a_{12}$, it is related to the structure multiplying $a_{17}$ by Eq.~\eqref{eq:axgor}.
\begin{align}
\bar{u}(p')\,\gamma^5 P^{\{\mu}\sigma^{\nu\}\alpha}\Delta_\alpha\,u(p) = 2iP^\mu P^\nu \bar{u}(p')\gamma^5 u(p).
\end{align}
Identical reasoning applies in relating the structures corresponding to $a_{14}$ and $a_{20}$.
In echo of $a_3$ the relation between $a_{13}$ to other terms is less immediate, but by employing the same strategy one finds
\begin{align}
\bar{u}(p')\,\gamma^5 \Delta^{\{\mu}\sigma^{\nu\}P}\,u(p) &= \frac{i\Delta^{\{\mu}}{2}\,\bar{u}(p')\,\gamma^5\Big(2m\gamma^{\nu\}} \nonumber \\
&+ \Delta^{\nu\}} \Big)u(p).
\end{align}
Therefore, $a_{13}$, $a_{16}$ and $a_{18}$ are linearly dependent.
The $a_{14}$ structure is related to $a_{15}$ and $a_{19}$ by the same line of reasoning.
\par
Therefore, eight of the terms in Eq.~\eqref{eq:overcomplete} can be absorbed into the other ones.
What remains from the original twenty is a set of twelve independent tensor structures, six of which correspond to factors independent of $\gamma_5$ and are present in the standard $\theta = 0$ case.
The other six are multiplied by $\gamma_5$ and therefore parity‑odd.
They are only present in the decomposition because the hadronic states are not eigenstates of parity.
To specify, we choose to retain the terms $a_1$, $a_3$, $a_5$, $a_6$, $a_7$, $a_{10}$ to align with familiar decompositions \cite{Ji:1996ek}.
The new correspondin terms kept are $a_{11}$, $a_{13}$, $a_{15}$, $a_{16}$, $a_{17}$, and $a_{20}$.
\par
Hermiticity of the EMT matrix element,
\begin{align} \label{eq:hermiticity}
    \langle p',s'|\,T^{\mu\nu}\,|p,s\rangle^\dagger = \langle p,s|\,T^{\mu\nu}\,|p',s'\rangle,
\end{align}
provides constraints on the real and imaginary components of the form factors.
Of the remaining terms that are parity-conserving $a_1$, $a_3$, $a_5$ and $a_{10}$ are clearly real-valued while $a_6$ and $a_7$ are purely imaginary.
When taking the conjugate of the terms containing $\gamma_5$, several factors of $(-1)$ arise from anticommuting with $\gamma_5$.
Therefore, in contrast to the previous case, the terms corresponding to $a_{11}$, $a_{13}$, $a_{16}$ and $a_{20}$ are purely imaginary, while the terms corresponding to $a_{15}$ and $a_{17}$ are real.
We introduce factors of $i$ into the tensor structures multiplying imaginary terms in order to keep all coefficient functions real-valued.
\par
Consider now the consequences of the discrete symmetries.
First note that while the $\theta$ term is separately parity- and time-reversal odd, it is C-invariant and therefore PT remains conserved.
Although parity and time reversal separately reverse the spatial momenta, their combined action leaves the four-momenta unchanged: $p^\mu\rightarrow p^\mu, \qquad p'^\mu\rightarrow p'^\mu, \qquad P^\mu\rightarrow P^\mu, \qquad \Delta^\mu\rightarrow\Delta^\mu.$
Spin, however, is flipped under $PT$.
So while factors of $P$ and $\Delta$ are unchanged, the external spinor labels are reversed.
In the following, $PT\left[ \right]$ denotes the transformation after applying the spinor transformation laws and expressing the result in the original spinor basis.
Because $PT$ is antiunitary, invariance requires 
\begin{align}
    PT \left[ \langle f | T^{\mu \nu} | i \rangle \right] =  \langle f | T^{\mu \nu} | i \rangle^*
\end{align}
The coefficient functions depend only on $t$, which remains unchanged under $PT$, and have been chosen real by hermiticity.
Consequently, any remaining spinor bilinear must transform into its complex conjugate.
Any component in the decomposition that instead transforms into minus its complex conjugate must have a vanishing coefficent.
Covariants that then transform into minus their complex conjugate must vanish.
Such minus signs will come from the transformation of the Dirac structures or the effect of commuting antiunitary operators through the factors of $i$ introduced to keep the coefficient functions real.
\par
As an explicit example, consider the $a_6$ term, 
\begin{align}
    \mathrm{PT} \Big[ \overline{u}(p') a_6(t) &i \gamma^{\{ \mu} \Delta^{\nu\}} u(p) \Big] \nonumber \\
    &= -i \overline{u}(p') a_6^*(t)  \gamma^{\{ \mu} \Delta^{\nu\}} u(p)
\end{align}
Since we know $a_6$ to be real, this combination is $PT$-odd requiring $a_6$ vanish.
A similar argument holds for $a_{10}$, therefore
\begin{align}
    a_6(t) = a_{10}(t) = 0.
\end{align}
Now consider the terms containing a $\gamma_5$ factor.
This time the tensor structures multiplying $a_{11}$, $a_{13}$, $a_{15}$, and $a_{16}$ are all odd under PT due to the additional $\gamma_5$ factor.
Whereas $a_{17}$ and $a_{20}$ are PT even.
However, only the terms $a_{11}$, $a_{13}$, $a_{16}$, and $a_{20}$ come with a compensating factor of $i$.
Therefore, 
\begin{align}
    a_{15}(t) = a_{20}(t) = 0.
\end{align}
Altogether, the constraints from hermiticity and PT invariance reduce the number of independent form factors to the eight shown in Eq.~\eqref{eq:emt-flavour-decomposition}. 
\par
The final reductions come from the conservation of the EMT, $\partial_\mu T^{\mu\nu} = 0$.
Applied to the matrix elements, this gives
\begin{align}
    \Delta_\mu \langle p',s'|\,T^{\mu\nu}\,|p,s\rangle = 0.
\end{align}
Contracting with the remaining structures gives the following relations between the form factors
\begin{align}
    0 &= \overline{u}(p') \Delta_{\mu} \Gamma^{\mu\nu} u(p) \nonumber \\
    &= \overline{u}(p') \Big[ \left(a_1 + t a_3 \right)  \Delta^\nu + i\gamma^5 \left(a_{11} + t a_{13} \right) \Delta^\nu   \nonumber \\
    &\hspace{ 1.5cm}+ \frac{i a_{16}}{2} \gamma_5\left( - 2 m \Delta^{\nu} + t \gamma^{\nu} \right) \Big] u(p)
\end{align}
where terms containing $P \cdot \Delta$ or $\slashed{\Delta}$ but not $\gamma_5$ have been removed since they vanish when contracted with spinors.
Gordon identities have also been used to re-express terms in a more familiar form.
This conservation equation must hold for every choice of external spins.
\par
For $t \neq 0$, the structures multiplying $a_{16}$ do not point along the same directions as the other terms in spin-space.
Then, for the conservation equation to hold, $ a_{16} $ must be 0.
The remaining scalar and pseudoscalar bilinears are also independent in the external spin space, as can be easily confirmed by evaluating the spinor bilinears in the Breit frame.
This means that their coefficients must vanish separately as well.
In total, there are three constraints
\begin{align} \label{}
    a_1(t) + t a_3(t) &= 0, \nonumber \\
    a_{11}(t) + t a_{13}(t) &= 0, \nonumber \\
    a_{16}(t) &= 0.
\end{align}
reducing the number of independent form factors to 5.
%
\section{Crossed-Channel $J^{PC}$ Counting}\label{App:jpc_count}
%
As an independent check of the decomposition above, we can count the number of allowed form factors using the crossed-channel argument of Ji and Lebed~\cite{Ji:2000id}, as reviewed for the EMT in regular QCD in Ref.~\cite{Cosyn:2019aio}.
Rather than explicitly constructing the tensor basis from the available Lorentz and Dirac structures, this method decomposes the rank-two operator into states of definite angular momentum and counts the number of corresponding hadron-antihadron partial waves.
The two approaches therefore enumerate the same space of tensor structures, but in different bases.
\par
Beginning with the off-forward matrix elements of the EMT, crossing symmetry relates these to the matrix elements of the EMT between the vacuum and a nucleon-antinucleon pair,
\begin{align}
    \langle p',\lambda'|\,T^{\mu\nu}\,|p,\lambda\rangle = \langle N(p',\lambda')\bar N(-p,\lambda)|\,T^{\mu\nu}\,|0\rangle.
\end{align}
In this channel, the momentum transfer of the elastic matrix element, $\Delta$, becomes the total momentum of the $N\bar N$ pair, while $P$ is related to their relative momentum.
The counting can then be performed in the center-of-mass frame of the crossed channel, where $\Delta^\mu = \left(\sqrt{t},\vec{0}\right)$.
Although this corresponds to a timelike value of $t$, the number of independent form factors is unchanged by analytic continuation.
\par
The general rank-two tensor transforms in the Lorentz representation as
\begin{align}
    \left(\frac{1}{2},\frac{1}{2}\right) \otimes \left(\frac{1}{2},\frac{1}{2}\right) \simeq (0,0)\oplus (1,0) \oplus (0,1) \oplus(1,1).
\end{align}
The $(1,0)$ and $(0,1)$ representations are antisymmetric and therefore do not contribute to the symmetric EMT considered here.
Therefore, the symmetric EMT transforms as $(0,0) \oplus (1,1)$.
Clearly, the $(0,0)$ representation contains a single scalar, $J=0$, component.
While the $(1,1)$ representation contains a single $J=0$ component, a three-component $J=1$ vector, and a five-component $J=2$ symmetric-traceless tensor,
\begin{align}
    (1,1) \simeq 0 \oplus 1 \oplus 2.
\end{align}
This decomposition is visible by inspecting the components of the symmetric EMT.
The component $T^{00}$ is a rotational scalar and therefore corresponds to $J=0$.
The three components $T^{0i}$ transform as a vector and correspond to $J=1$.
Finally, the spatial tensor $T^{ij}$ can be separated into its trace and symmetric-traceless pieces,
\begin{align}
    T^{ij}
    =
    \frac{1}{3}\delta^{ij}T^{kk}
    +
    \left(
    T^{ij}-\frac{1}{3}\delta^{ij}T^{kk}
    \right),
\end{align}
which correspond to $J=0$ and $J=2$, respectively.
Therefore, before imposing conservation, the symmetric EMT contains two $J=0$ components, one $J=1$ component, and one $J=2$ component.
The dimensions provide a simple check,
\begin{align} \label{eq:Jdecomposition}
    2(1)+3+5=10,
\end{align}
which is the number of independent components of a symmetric rank-two tensor in four dimensions.
\par
For the usual parity-conserving case, these components may be labelled by
\begin{align}
    (0,0)&:\qquad 0^{++}, \nonumber\\
    (1,1)&:\qquad 0^{++},\ 1^{-+},\ 2^{++}.
\end{align}
The second superscript is fixed because the EMT is even under charge conjugation.
For the case considered here, however, parity is not a symmetry of the theory.
The $J^{PC}$ labels remain useful for identifying the transformation properties of the possible $N\bar N$ partial waves, but parity can no longer be imposed as a selection rule.
Charge conjugation remains a symmetry of QCD with a $\theta$ term, so the crossed-channel state must still have $C=+$.
Thus, for each angular momentum $J$ appearing in Eq.~\eqref{eq:Jdecomposition}, we must count all $N\bar N$ partial waves with the same $J$ and $C=+$, irrespective of their parity.
\par
For a fermion-antifermion pair, the parity and charge-conjugation quantum numbers are
\begin{align}
    P &= (-1)^{L+1},\\
    C &= (-1)^{L+S},
\end{align}
where $L$ and $S$ are the orbital and total spin angular momenta, respectively.
The allowed $C$-even partial waves for $J\leq 2$ are given in Table~\ref{tab:JPCcounting}.
\begin{table}[ht]
    \centering
    \renewcommand{\arraystretch}{1.3}
    \begin{tabular}{cc}
        \toprule
        \textbf{$J$} & \textbf{$N\bar N$ partial waves ($L$)}  \\
        \midrule
        $0$ & $0^{-+}(0),\quad 0^{++}(1)$  \\
        $1$ & $1^{++}(1)$ \\
        $2$ & $2^{-+}(2),\quad 2^{++}(1),\quad 2^{++}(3)$  \\
        \bottomrule
    \end{tabular}
    \caption{The $C$-even spin-$1/2$ hadron-antihadron partial waves which can couple to the angular-momentum components of a symmetric rank-two tensor. The orbtial quantum number is displayed for classification.}
    \label{tab:JPCcounting}
\end{table}
There are two independent $J=0$ components of the symmetric tensor, and each can couple to either of the two $J=0$, $C$-even partial waves.
These therefore give four independent amplitudes.
Since parity is no longer a symmetry of the theory, the single $J=1$ component can couple to one $C$-even partial wave, while the single $J=2$ component can couple to three.
The total number of independent form factors before imposing conservation is therefore
\begin{align}
    N_{\mathrm{GFF}} = \left( 2\times 2 \right) + \left(1\times 1\right) + \left(1\times 3\right) = 8.
    \label{eq:JPCeight}
\end{align}
This agrees with the eight independent structures obtained from the explicit Lorentz and Dirac decomposition above after imposing hermiticity and the surviving discrete symmetries.
\par
As a check, consider restoring parity invariance.
The two $J=0$ components of the EMT both have positive parity and can therefore couple only to the $0^{++}$ channel.
The $J=1$ component has quantum numbers $1^{-+}$, for which there is no corresponding spin-$1/2$ $N\bar N$ partial wave.
Finally, the $2^{++}$ component can couple to the ${}^{3}P_2$ and ${}^{3}F_2$ channels.
The count then becomes
\begin{align}
    N_{\mathrm{GFF}}^{P\text{-even}} = 2+0+2 = 4,
\end{align}
which is the familiar number of form factors for the symmetric EMT, without conservation.
The additional four amplitudes in Eq.~\eqref{eq:JPCeight} are therefore precisely those made available by relaxing parity invariance.
\par
We can now impose conservation directly within the crossed-channel counting.
For the total EMT,
\begin{align}
    \Delta_\mu \langle N\bar N|\,T^{\mu\nu}\,|0\rangle = 0.
\end{align}
In the center-of-mass frame, $\Delta^\mu=(\sqrt{t},\vec{0})$, so for generic non-zero $t$ this gives
\begin{align}
    \bra{N \bar{N} }T^{0\nu}\ket{0}=0.
\end{align}
Consequently, the $T^{00}$ component and all of the $T^{0i}$ components are removed.
In terms of angular momentum, conservation therefore removes one of the two $J=0$ components and the complete $J=1$ component.
The remaining $J=0$ component can couple to the two channels $0^{-+}(L=0)$ and $0^{++}(1)$, while the $J=2$ component can couple to $2^{-+}(2)$, $2^{++}(1)$, and $2^{++}(3)$.
Therefore, the conserved symmetric EMT contains $N_{\mathrm{GFF}}^{\mathrm{cons.}} = 2+3 = 5$ independent form factors.
Of these five amplitudes, the positive-parity channels $0^{++}(1)$, $2^{++}(1)$, and $2^{++}(3)$ account for the three standard GFFs, and the negative-parity channels $0^{-+}(0)$ and $2^{-+}(2)$ account for the two additional structures allowed when parity is violated.
Thus, the crossed-channel counting independently reproduces the result of the explicit covariant decomposition.
\par
This also provides a useful interpretation of the three constraints obtained from EMT conservation in the previous section.
Conservation removes one parity-even $J=0$ amplitude, one parity-odd $J=0$ amplitude, and the single $C$-even $J=1$ amplitude.
This agrees with the three relations
\begin{align}
    a_1(t)+t a_3(t) &= 0, \nonumber\\
    a_{11}(t)+t a_{13}(t) &= 0, \nonumber\\
    a_{16}(t) &= 0,
\end{align}
obtained directly from the Lorentz decomposition.
The $J^{PC}$ counting provides a representation-theoretic enumeration of the same independent matrix elements and a useful check that the explicit tensor basis is both complete and minimal.
\bibliography{References}

\end{document}